\documentclass[a4paper]{jpconf}  % for double-spaced preprint
\usepackage[dvips]{graphicx}
\usepackage{dcolumn}   % needed for some tables
\usepackage{bm}        % for math
\usepackage{amssymb}   % for math
\usepackage[utf8]{inputenc}
\usepackage{mathrsfs}
\usepackage{amsmath}
\usepackage{hyperref}
\usepackage{textcomp}
\usepackage{subfigure}
\usepackage{color}
\usepackage{cite}

\newcommand{\gsim}{\stackrel{\scriptstyle >}{\phantom{}_{\sim}}}

\begin{document}

\title{Superheavy objects composed of nuclear and dark matter}

\author{Vakhid A. Gani$^{1,2}$, Maxim Yu. Khlopov$^{1,3}$, Dmitry N. Voskresensky$^{1,4}$ }

\address{$^1$National Research Nuclear University MEPhI (Moscow Engineering Physics Institute), 115409 Moscow, Russia}
\address{$^2$National Research Center Kurchatov Institute, Institute for Theoretical and Experimental Physics, 117218 Moscow, Russia}
\address{$^3$Institute of Physics, Southern Federal University, 344090 Rostov on Don, Russia}
\address{$^4$Joint Institute for Nuclear Research, 141980 Dubna, Moscow region, Russia}

\ead{vagani@mephi.ru}
%\begin{frontmatter}
\begin{abstract}

We consider a model of $O$He atomic dark matter formed by Coulomb binding of the hypothetical stable double-charged massive $O^{--}$  particles with nuclei of primordial helium. Such a dark matter can be captured by ordinary matter forming superheavy nuclei. We propose a possibility of self-bound by nuclear and electromagnetic interaction $O$-nuclearites as well as more massive gravitating ones and discuss effect of accumulation of $O$He atoms in stars and their effect in stellar evolution.

\end{abstract}%\end{frontmatter}

%%%%%%%%%%%%%%%%%%%%%%%%%%%%%%%%%%%%%%%%

\section{Introduction}
The nature of cosmological dark matter (DM) is related to physics beyond the Standard model. Models of dark $O$He atoms formed by hypothetical $-2$ charged particles, which are bound with nuclei of primordial helium, occupy special place in the list of DM candidates. In such models the only free parameter is the mass of $-2$ charged electromagnetically interacting massive particle and many features of this type of DM can be described by the known nuclear and atomic physics. Thereby such a description is simplest from the point of view of new physics.

Several models predict stable double charged particles in absence of stable single charged particles. Appropriate candidate is the hypothetical heavy stable quark $U$ of fourth generation  \cite{Glashow:2005jy}. If an excess of $(-2/3)$ charged antiquarks $\bar{U}$ is generated in the early Universe, then excessive antiquarks form $-2$ charged antibaryons $\bar{U}\bar{U}\bar{U}$. The latter are captured by primordial helium thus forming $O^{--}$He$^{++}$ ($O$He) atoms \cite{Khlopov:2005ew}. This hypothesis was extensively discussed in the literature, see \cite{Khlopov:2010ik,gKhlopov,khlopov.ijmpd.2015,Wallemacq:2015cjr} and references therein. The model can explain the observed excess of positronium annihilation line in the galactic bulge and excessive fraction of high energy cosmic ray positrons (in case the mass of $O$-particle doesn't exceed 1.3 TeV), thus challenging direct test of this hypothesis in searches for stable double charged particles at the Large Hadron Collider \cite{probes}.

On the other hand, various hypotheses of existence of stable superheavy  nuclei and nuclearites glued by pion condensate \cite{Migdal:1971cu,Migdal:1977rn,Voskresensky:1977mz,Migdal:1978az,Migdal:1990vm}, sigma condensate \cite{Lee:1974kn}, either by strange quarks \cite{Bodmer:1971we,Witten:1984rs,DeRujula:1984axn,Alcock:1986hz,DeRujula:1989fe} have been explored. These objects may exist in the Galaxy as debris from the Big Bang and various astrophysical catastrophes, e.g., supernovae explosions, phase transitions in neutron stars and star collisions \cite{Migdal:1990vm,DeRujula:1984axn,Weber}. Another important issue is a possible influence of DM captured by stars on their structure and evolution may lead to observable effects, e.g., in compact stars \cite{Kouvaris.PRD.2014}.

In this contribution we argue that colliding with nuclei of ordinary matter $O$He atoms may form superheavy $O$-enriched nuclei. Further we explore a possible existence of $O$-nuclearites constructed of nuclear matter with density typical for the nuclear saturation, where negative electric charge of $O^{--}$ and positive electric charge of protons are compensated. Then we study effect  of such nuclearites at astrophysical conditions. A more detailed consideration can be found in \cite{GKV2018}.

%%%%%%%%%%%%%%%%%%%%%%%%%%%%%%%%%%%%%%%%

\section{Heavy nuclei consisting $O$-DM particles  and self-bound $O$-nuclearites}

Consider a rather heavy isospin-symmetric nucleus of atomic number $A=2N_\mathrm{p}$ (the number of neutrons equals the number of protons, $N_\mathrm{n}=N_\mathrm{p}$).  Assume that heavy $O$-particles entering the nucleus are distributed there with density $n_O (r)$. The energy of this $O$-nucleus is
\begin{equation}\label{E}
{\cal{E}}=-16 {\rm{
 MeV}} \cdot A  - \int d^3 r (n_\mathrm{p} -2n_O)V -\int d^3 r\frac{(\nabla{V})^2}{8\pi e^2} + {\cal{E}}_{\rm kin}^{O}\, ,
\end{equation}
where the first term is the atomic nucleus volume energy, while the second and third terms yield the electromagnetic energy \cite{Migdal:1977rn,Migdal:1990vm}, and
\begin{equation}\label{E_kin}
{\cal{E}}_{\rm kin}^O = \displaystyle\int d^3 r \int\limits_0^{p_{{\rm F},O}^{}} \frac{p^2 dp}{\pi^2}\frac{p^2}{2m_O^{}}
\end{equation}
is the kinetic energy of the $O$-fermions of mass $m_O$; $V = -e\phi$ is the potential well for the protons in the field of the negatively charged $O$-particles, $n_O^{}=p_{{\rm F},O}^3/(3\pi^2)$, $n_{\rm p} = n_\mathrm{n} = p_{{\rm F},p}^3/(3\pi^2)$.
Following \cite{Khlopov:2005ew} in our estimations we assume $m_O\simeq $TeV, i.e.\ $m_O\gg m_{\rm N}$, where $m_{\rm N}$ is the nucleon mass. Hence the contribution of ${\cal{E}}_{\rm kin}^{O}$ is tiny, and can be neglected. Besides we neglect a small nuclear surface term compared to the volume one.

The most energetically favorable distribution of negatively charged $O$-particles inside the nucleus is that follows the proton one. Except a narrow nuclear diffuseness layer $\delta R\sim 0.5$ fm near the surface, the proton and neutron number densities in  electrically neutral nucleus of a large size  are $n_\mathrm{p} = n_\mathrm{n} \simeq  n_\mathrm{p}^0 \theta(r-R)$, where $2n_\mathrm{p}^0 = n_0 = 0.16$ fm$^{-3}$ is the normal nuclear saturation density. Thereby, if the number of $O$-particles $N_O\geq A/4$, their final density inside the  nucleus becomes $n_O = n_\mathrm{p}/2 = (n_\mathrm{p}^0/2)\:\theta(r-R)$, excessive $O$-particles are pushed out. Thus constructed $O$-superheavy nucleus (for $A\gsim 100$) and $O$-nuclearite (for $A\gg 100$) are absolutely stable and have the energy ${\cal{E}}\simeq -16 {\rm{MeV}} \cdot A$ for arbitrary $A$. If antimatter existed in sufficient amount, on equal footing we could consider anti-superheavy nuclei and nuclearites made of $\bar{p}$ and $\bar{n}$ with the electric charge compensated by $O^{++}=UUU$.

We further assume that the mass-density $\rho_\mathrm{DM}^{} \simeq \rho_{O{\rm He}}^{}$ (the local density of a non-luminous mass in the Galaxies is $\rho_\mathrm{DM}^{} \simeq (3-7)\cdot 10^{-25}$g$/$cm$^3$, cf.\ \cite{Chin:1979yb}) and that $O$He interacts dominantly elastic with ordinary matter. Nevertheless, if $O$-particle enters a nucleus of ordinary matter, then it is energetically profitable for it to remain there, thus making the nucleus superheavy. Ordinary nuclei, passing through a permanent $O$He flux, with some probability absorb $O$- and $\alpha$-particles producing new $\{OA\}_{N_\mathrm{n}}^{N_\mathrm{p}}$ nuclei.  Since $O$-nucleus has not been observed yet, such events should be very rare.

%%%%%%%%%%%%%%%%%%%%%%%%%%%%%%%%%%%%%%%%

\section{Gravitating $O$-nuclearites and black holes}

 Above we considered a self-bound nuclearite with $n=n_0$ in absence of gravity. However for  very large $A$ the gravity comes into play.
For an individual  $O$-nuclearite to remain in a self-bound state the nucleon density should be $n< (2-2.5) n_0$, since for such densities the energy of iso-symmetric nuclear matter (at switched off the Coulomb term) remains negative, cf.\ \cite{Klahn:2006ir}.

Here we roughly estimate size of the gravitationally-stable $O$-nuclearite. In a stable object the internal (nucleon) pressure should be compensated by the gravitational one. Assume that ${\cal{E}}_{\rm kin}^{\rm nucl}\ge |{\cal{E}}_{\rm pot}^{\rm nucl}|$, then the internal pressure is mainly determined by the Fermi gas of the nucleons, ${\cal{E}}_{\rm kin}^{\rm nucl}\sim (p_{{\rm F}, \rm nucl}^2/m_\mathrm{N}) A$, $p_{{\rm F},\rm nucl}\sim A^{1/3}/R$. The gravitational energy is ${\cal{E}}_{\rm grav}\sim -GM^2/R$, $M\simeq A m_O$. Thus minimizing in $R$ we  estimate the size of the object
\begin{eqnarray}\label{R}
R\sim 1/(GM^{1/3}m_\mathrm{N}^{} m_O^{5/3})\sim 10 {\rm{km}} (M_{\odot}/M)^{1/3}(m_\mathrm{N}^{}/m_O)^{5/3}\,.
\end{eqnarray}
As follows from (\ref{R}) an individual $O$-nuclearite has central density $n\sim (2-2.5) n_0$, if its mass $M\sim 3\cdot 10^{-8}M_{\odot}$ and the radius $R\sim 30$ m (for $m_O^{}\simeq 10^3 m_\mathrm{N}$). Thereby a neutron star of the mass $M\ge M_{\odot}$ may accumulate at most several times $10^7$ of most heavy self-bound nuclearites before it collapses to a black hole.

%%%%%%%%%%%%%%%%%%%%%%%%%%%%%%%%%%%%%%%%

\section{Accumulation of $O$-nuclearites during the star evolution}

Consider accretion of $O$He flux onto a neutron star. Self-bound $O$-nuclearites might be formed in the central regions of neutron stars with masses corresponding to central densities $n_{\rm cen}\le (2-2.5) n_0$. $O$He dissociates already not far from the crust-core boundary (for $n\gsim n_0$), and the $O$-particles dive down towards the center of the neutron star.

The number of $O$-particles in a neutron star, $N_O$, obeys equation $dN_O/dt=C_{\rm capt}$, where $C_{\rm capt}$ is the rate of $O$He capture through scattering by baryons. The capture can occur only if the transferred momentum is larger than the difference between the Fermi momentum of the baryon and the momentum of the re-scattered baryon. For $m_O^{}\gg m_\mathrm{N}^{}$ we obtain \cite{Gould:1987ju,Gould:1987ir,McDermott:2011jp}
\begin{eqnarray}\label{escape}
\frac{dC_{\rm cap}}{d\Omega_3}\simeq \sum_b \sqrt{\frac{6}{\pi}}\frac{\rho_{O{\rm He}}^{}(r)}{m_O^{}}\frac{v^2(r)}{\bar{v}^2}n_b (r)(\bar{v}\sigma_{O{\rm He},b})\xi_b\left[ 1-\frac{1-e^{-B_b^2}}{B_b^2} \right]\,,
\end{eqnarray}
where $\Omega_3$ is the volume of the neutron  star, $\rho_{O{\rm He}}^{}(r)$ is the ambient $O{\rm He}$ mass density, $n_b (r)$ is the number density of the baryon species $b=(n,p, H,...)$, $H= \Lambda ,\Sigma,\Xi$, $v(r)$ is the escape velocity of the neutron star at the radius $r$, $\bar{v}$ is the $O$He-velocity dispersion around the neutron star, $\sigma_{O{\rm He},b}$ is the effective scattering cross section between $O$He and baryon $b$ in the neutron star, $\xi_b=\mbox{min}\{\delta p_b /p_{{\rm F},b},1\}$ takes into account the effect of the neutron degeneracy on the capture, $\delta p_b\simeq \sqrt{2} m_{\rm red}v_{\rm esc}$, $m_{\rm red}$ is the reduced $O$He--baryon mass, $m_{\rm red}\simeq m_\mathrm{N}^{}$, $p_{{\rm F},b}^{}$ is the $b$-baryon Fermi-momentum, $B_b^2 \simeq 6 m_b v^2(r)/(m_O^{} \bar{v}^2)$.

Near the boundary of the neutron-star crust--core $n\sim n_0$ and $n_\mathrm{n}\gg n_\mathrm{p}$, $n_H^{}=0$. Typically \cite{McDermott:2011jp} $v_{\rm esc}\sim v(r\sim R)\sim p_{{\rm F},n}^{}/m_\mathrm{N}^{}\sim 10^5$ km$/$s for $n\simeq n_\mathrm{n}\sim n_0$ and thus $\xi_b \sim 1$, $\bar{v}\simeq 250$ km$/$s, hence $B_b\gg 1$. Then equation \eqref{escape} yields $C_{\rm cap} \sim \displaystyle\frac{\rho_{O{\rm He}}}{m_{O}}\frac{v_{\rm esc}^2}{\bar{v}^2}\bar{v}\sigma_{O{\rm He},n}N_n$, and $N_O\simeq C_{\rm cap} t$.

The maximum value of $\sigma_{O{\rm He},n}^{}$ is $\pi R^2/N_\mathrm{n}$, and we can estimate the maximum number of $N_O^{\rm max}$ and the maximum mass of $O$-nuclearites which are accumulated in the center of a neutron star of the given age $N_O^{\rm max}\sim 10^{39}\frac{t}{10^{10}{\rm yr}}$, $M_{O{\rm-nuclearite}}^{\rm max}\sim 10^{18}\frac{t}{10^{10}{\rm yr}}{\rm g}$. The radius of a self-bound $O$-nuclearite is found from equality $n\,\Omega_{O{\rm-nuclearite}}=N_O$, and we obtain $R^{\rm max}\sim {\rm cm}$. Thus in order to accumulate inside the old (of the age $\sim 10^{10}$yr.) neutron star a mass $M_{O{\rm-nuclearite}}\sim M_{\odot}$, $O$He flux onto the neutron star should be $\sim 10^7$ times enhanced compared to that we used in the above estimates.

Red giants also can accumulate $O$He matter in the star centers during their evolution. Using $R\sim 10^9 $ km, $M\sim 0.5 M_{\odot}$, $t_{\rm life}\sim 10^8$yr., $v_{\rm esc}\sim \bar{v}$ we have $N_O^{\rm max}\sim 10^{46}$ and $M_{O{\rm -nuclearite}}^{\rm max}\sim 10^{25}$ g. Similar estimates are also valid for red supergiants. The $O$He nugget, being formed in the star center, awaits the supernova explosion. When nucleons begin to fall down to the center, a self-bound $O$-nuclearite can be formed.

%%%%%%%%%%%%%%%%%%%%%%%%%%%%%%%%%%%%%%%%

\section{Conclusion}

Hypothesis of dark atoms of $O$He is of special interest due to minimal involvement of new physics in their properties. Indirect astrophysical effects of $O$He DM are related to a radiation from $O$He excitation in the collisions in the Galactic Center. It may give an explanation of excess of positronium annihilation line, observed by INTEGRAL in the galactic bulge, if the mass of $O^{--}$ particle is near $1.25$ TeV.

Interplay between elastic and inelastic interactions of $O$He atoms with ordinary matter   remains  open problem. Crucial point here  is  existence of a potential barrier in the interaction of $O$He with ordinary  nuclei. Putting aside this problem we  studied here some consequences  of the capture of $O$He atoms by  nuclei. We proposed a possibility of existence of stable superheavy nuclei,  self-bound $O$-nuclearites and gravitating more massive ones and we discussed various mechanisms for their accumulation in stars.

%%%%%%%%%%%%%%%%%%%%%%%%%%%%%%%%%%%%%%%%

\section{Acknowledgments}

The work of V.A.G.\ on studies of $O$-nuclearites was supported by the MEPhI Academic Excellence Project (contract No.\ 02.a03.21.0005, 27.08.2013). The work of M.Yu.K.\ on studies of effects of $O$He dark matter was supported by grant of the Russian Science Foundation (project No-18-12-00213). The work of D.N.V.\ on studies of the charged massive particle binding in $O$-nuclearites was also supported by the Ministry of Education and Science of the Russian Federation within the state assignment, project No~3.6062.2017/6.7.

\section*{References}

% For arXiv:

\vspace{-5.7mm}

\begin{figure}[h!]
\centering
\includegraphics[width=0.10\textwidth]{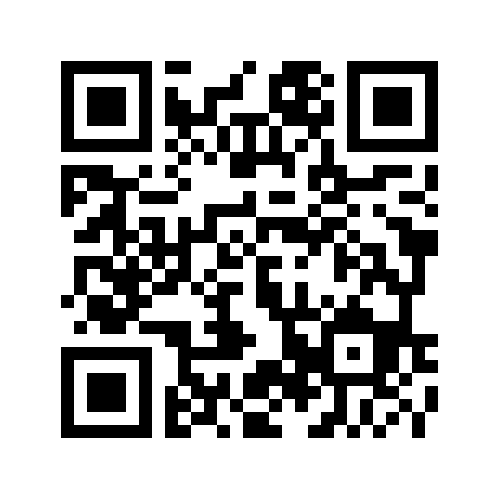}
\end{figure}

\end{document}